\documentclass[intlimits,twoside,a4paper]{article}

\usepackage[eqsecnum]{cmpj3}

\issue{2024}{27}{1}{13602}
\doinumber{10.5488/CMP.27.13602}

\title[{Shape changes of a single hairy particle with mobile ligands at a liquid-liquid interface}]%
{Shape changes of a single hairy particle with mobile ligands at a liquid-liquid interface}
\author[T. Staszewski, M. Bor\'owko]{T. Staszewski\orcid{0000-0002-0284-4253}, M. Bor\'owko\orcid{0000-0003-1461-249X}}
\address{Department of Theoretical Chemistry, Institute of Chemical Sciences, Faculty of Chemistry, Maria Curie-Sk{\l}odowska University in Lublin, Poland}
\Keywords{hairy particles, particle-laden layers, molecular dynamics}

\date{Received July 5, 2023, in final form November 8, 2023}

\begin{document}

\maketitle

\begin{abstract}
{\color{black}
We investigate rearrangements of a single hairy particle at a liquid-liquid interface using coarse-grained molecular
dynamics simulations. We consider the particles with the same (symmetrical interactions) and different (asymmetrical
interactions) affinity to the liquids. We show how ligand mobility affects the behavior of the hairy particle
at the liquid-liquid interface. We found that such a hairy particle can take various shapes at the interface. For
example, a Janus-like snowman consisting of a segment cluster and a bare part of the core, Saturn-like structures,
and the core with a wide ``plume'' on one side. A configuration of the particle at the interface is characterized
by the vertical displacement distance and the orientation of the particle relative to the phase boundary. The selected
descriptors are used to characterize the shape of the segment cloud. We found that the shape of a particle and its localization at the interface
can be determined by tuning the interactions with the liquids.}

%
\printkeywords
%
\end{abstract}

\section{Introduction}

Polymer-tethered (hairy) nanoparticles have received much attention in nanomaterials science {\color{black} \cite{2,3,4}}.   They are composed of a core and a layer of polymer chains grafted by one end on the core's surface.  Such hybrid particles naturally combine the features of both ligands and cores.  Their properties can be tuned by altering the chemical nature, grafting density, chain length, and grafting architecture of the polymer layer, or by changing the nanoparticle size and shape.  Due to their unique properties, hairy particles have a broad range of applications, such as stabilization of  Pickering emulsions, production of composites, sensors, drug delivery systems, pollution removal, etc. {\color{black} \cite{2,3,4}}. 

{\color{black} Most of the research focused on the modelling of the morphology of polymer layers.  To solve this problem various theoretical approaches were used \cite{8,9,10,11,12,13}.} Ohno et al. \cite{8} proposed the mean-field theory of the polymer-tethered spherical particles of different sizes which is an extension of the Daud-Cotton method developed for star polymers \cite{9}. The self-consistent field model and the scaling theory were also used to study configurations of chains tethered on spherical particles \cite{10,11}. Moreover,  Lo Verso et al. \cite{12} used the density functional theory to study polymers end-grafted to spherical nanoparticles under good solvent conditions. However, Ginzburg \cite{13} used a self-consistent field-density functional theory to investigate the self-assembly of hairy particles. The internal structure of hairy particles was also studied using molecular simulations {\color{black}\cite{14,15,16,18,19,20,21,22}}. Dong and Zhou \cite{14} investigated the responsive behavior of hairy particles in different solvents. They considered the particles modified with block copolymers or mixed polymer brushes and found various structures depending on the type of ligands and solvents.  The reconfiguration inside the polymer canopy can result in the formation of patchy nanoparticles~\cite{19,20}.  Molecular simulations also showed that the adsorption of small particles on ligands can change the structure of the polymer shell \cite{21,22}.  
{\color{black} Moreover, the variability of the internal structure of ligand-tethered particles considerably affects their assembly. Numerous studies have focused on the self-assembly of hairy particles in bulk systems, as these particles are promising building blocks for the production of novel materials \cite{28,29,30,31,32}.}


{\color{black}All theoretical studies have clearly shown that in bulk systems hairy particles change their shapes in response to the environment. Their shape-flexibility depends on the nature of ligands, solvents, grafting density, and temperature. These results are a signpost for studying hairy particles at fluid-solid and fluid-fluid interfaces.}
 
{\color{black}The materials containing nanoparticles can release them into the environment. Therefore, it is necessary to develop the methods of removing such pollutants. The adsorption on solid surfaces was considered the most effective and safe procedure for the removal of nanoparticles \cite{4}. The amount of adsorption on solids depends on the size and shape of the particles. Therefore, changes in the shape of hairy particles on the solid surfaces  and inside the pores were  investigated \cite{r1, r2, 24,25,26}.}

{\color{black}The behavior of the hairy particles at fluid-fluid interfaces was also extensively studied both from experimental and theoretical perspectives \cite{33,34,35,36,37,38,39}. Numerous medical applications of hairy particles require a deep understanding of the mechanism of their attachment at interfaces and the transfer to cells~\cite{3}.  It has been well established that surface-active moieties (e.g., surfactant, (bio)polymers, and micro- and nanoparticles) tend to accumulate at fluid-fluid interfaces. The adsorption is driven by minimization of the Gibbs free energy owing to a decrease in the contact area between the two fluids. The energies associated with the attachment of particles to fluid interfaces depend on the chemical nature, size, roughness, and wettability of the particles characterized by their contact angles \cite{r3, 33, 34}. Due to large sizes of nanoparticles, their attatchement enegies  at fluid-fluid interfaces are much higher than those of standard amphiphiles. For this reason, the adsorption of the particles at the droplet-continuous phase interface is essential for the stabilization of these systems  \cite{4, r3, r4}. The properties of hairy particles can be tuned in a wide range by changing the nature of polymer coating. Thus, they can be considered as potentially highly efficient emulsifiers.} 
{\color{black} The morphology of polymer coatings plays a significant role in these processes. At liquid-liquid interfaces, the chains on the two sides of the interface can adopt different
configurations depending on the nature of the ligands, the quality of both solvents, the grafting density, and other parameters \cite{34}. }

The structure of individual hairy particles adsorbed at the interface was extensively investigated by molecular simulations \cite{36,37,38,39}.
Tay and Bresme simulated the alkylthiol passivated gold nanocrystals trapped at the air-water interface and showed that the shape of the hairy particles was strongly perturbed by the interface \cite{35}.  Quan et al. \cite{37} studied the internal structure of gold nanoparticles modified with amphiphilic brushes at the oil-water interface. They discussed the influence of grafting architecture (diblock, mixed, and Janus brush-grafted particles) and the hydrophilicity of polymers on particle morphology.
Quite recently, Tang et al. \cite{38} studied the hairy particles on the interface between two mutually
immiscible polymers. They concluded that the free energy of rearrangement
of the ligands significantly contributes to the total change in free energy upon adsorption
of the particle.  The Brownian diffusion of various hairy particles at the liquid-liquid interface was also simulated and a good agreement between experimental and simulation results was obtained only when the flexibility of the particle shape had been taken into account \cite{39}. 

{\color{black} The analysis of the behavior of individual hairy particles at liquid-liquid phase boundaries can be a starting point for the study of their self-assembly.  Hairy particles can form either disordered monolayers or different two-dimensional structures of a high degree of ordering at these interfaces \cite{4, 33, 34}. Menath et. al~\cite{r7} proved that the changes in the shape of hairy particles at the interface considerably influence their self-assembly. }

{\color{black}As has been shown above, the capability of hairy particles to change their shapes in various environments is important for numerous processes. Although the influence of many factors on the morphology of the polymer coatings of hairy particles has been assessed, the mobility of ligands has rarely been taken into account.  It is well known, however, that the properties of hairy particles depend on the type of grafting bonds \cite{34,48}. For example, the thiol-gold bond is mobile, allowing thiolated ligands to move along the nanoparticle surface, while polyelectrolyte brushes grown from the surface of silica particles are irreversibly attached \cite{34}. Most theoretical models assumed that the chains were permanently anchored at certain points on the core surface
\cite{8,9,10,11,12,13,14,15,16,18,19}. Only a few theoretical works have dealt with mobile ligands~\cite{26,28,r5}. In such approaches, the binding segments of ligands can freely move on the core surface but they are incapable of breaking away from it. Ligand mobility can significantly influence the behavior of hairy particles.  For example, under certain conditions, the chains can form clumps on the core~\cite{19,20}. This, in turn, determines the self-assembly of hairy particles. Molecular simulations showed that hairy particles with fixed and mobile ligands can assemble in completely different structures \cite{28}.}

{\color{black}In this work, we use molecular dynamics to study individual hairy particles at liquid-liquid interfaces. 
Our study is motivated by the well-known phenomenon of nanoparticles migrating to liquid-liquid interfaces to reduce their surface energy. We take advantage of this effect to trap hairy particles at the interface
between two mutually immiscible liquids. The particles that interact similarly with both liquids  would be expected to localize at the interfacial plane, where they can occlude the largest possible area of the
interface and thereby provide the largest free energy benefit. As a consequence, the particle adsorption lowers the surface tension and stabilizes the interface  \cite{4,r3,r4,33,34}. Nanoparticles are strong emulsifiers~\cite{r4}.
In the case of particles that differently interact with the liquids forming the phase boundary,  these effects are stronger. 
As it has been mentioned, the ability of nanoparticles to stabilize the emulsions depends on the area of the occupied interface \cite{4,r3,r4}. Therefore, we discuss here the shape transformations of hairy particles at the liquid-liquid interface. We consider the particles with mobile ligands. In the case of such particles, changes in shape, and thus in the area of the occupied phase boundary, should be more visible. The effects of ligand mobility on the behavior of hairy particles at liquid-liquid interfaces have not been investigated so far.}

The rest of the work is organized as follows. In section~\ref{sec2}, we describe the model used and the simulation method and define the calculated observables.  Our results are presented and discussed in section~\ref{sec3}.   Finally, in section~\ref{sec4} we summarize and draw our conclusions.

\section{Methods}\label{sec2}
\subsection{Model}

We consider a coarse-grained model of the system that contains two immiscible fluids, $W$, $O$, and hairy particles. A single hairy particle is modelled as a spherical core with attached linear chains. The polymer chains, are treated as Kremer-Grest bead-chains \cite{48}, in which polymer segments, typically a few monomers long, are represented by beads of size $\sigma_S$ and mass $m_S$. We assume that all ligands have the same length (each chain consists of $M$ segments) and the number of ligands attached to each core is $f$. The cores are treated as soft spheres of diameter $\sigma_C$ and mass $m_C$. To {\color{black}reduce} the number of parameters, we assume that all segments and molecules of both fluids have the same diameters and masses.

Adjacent beads along the chain representing bonded segments are connected to each other using a finitely extensible nonlinear elastic (FENE) segment-segment potential

\begin{equation} \label{eq1}
U_{\text{FENE}}=-\frac{k}{2}R_0^2 \ln \left[ 1- \left( \frac{r}{R_0} \right)^2 \right],
\end{equation} 
where $r$ is the separation distance between the segments, $k$ is the spring constant and $R_0$ is the maximum possible length of the spring. {\color{black} We introduce a typical value  of  the  constant $k=30$ \cite{48}. For the chain backbone, we assume that $R_0=1.5\sigma_S$.

The same potential is used to model the binding between the core and the first segments of the ligands. In this case, however,  $R_0=1.5\sigma_{CS}$, where $\sigma_{CS}=0.5(\sigma_{C}+\sigma_{S})$. This potential allows the ligands to slide along the surface of the core.}

The interactions between nonbonded segments, liquid molecules, and cores are treated using  a cut-and-shifted Lennard-Jones potential \cite{49}

\begin{equation} \label{eq2}
u^{(ij)}=\left\{
\begin{array}{ll}
4 \varepsilon_{ij} \left[ (\sigma_{ij}/r)^{12}-(\sigma_{ij}/r)^6\ \right] + \Delta u^{(ij)}(r), & \ \ \ r < r_{\text{cut}}^{(ij)}, \\
0, & \ \ \ {\rm otherwise,}
\end{array}
\right.
\end{equation}
where
\begin{equation} \label{eq3}
\Delta u^{(ij)}(r)=-(r-r_{\text{cut}}^{(ij)})\partial u^{(ij)}(r_{\text{cut}}^{(ij)}) / \partial r,
\end{equation} 
and $r^{(ij)}_{\text{cut}}$ is the cutoff distance, $r^{(ij)}_{\text{cut}}=0.5(\sigma_i+\sigma_j)$ $(i,j = C,S,O,W)$,  $\varepsilon_{ij}$ denotes the parameter that characterizes interaction strengths between spherical species $i$ and $j$. The indices $C, S, W, O$ correspond to the cores, segments, liquid $W$, and  liquid $O$, respectively. We use the cutoff distance to switch on or switch off the attractive interactions. For the attractive  interactions, $r^{(ij)}_{\text{cut}}=2.5\sigma_{(ij)}$  while, for repulsive interactions, $r^{(ij)}_{\text{cut}}=\sigma_{(ij)}$.

Standard units are used herein. The diameter of segments is the distance unit, $\sigma=\sigma_S$ while the segments-segment energy parameter, $\varepsilon=\varepsilon_{SS}$ is the energy unit. The mass of a single segment is the mass {\color{black}unit}, $m_S=m$. The basic unit of time is $\tau=\sigma(\varepsilon/m)^{1/2}$. The gravity effect is assumed to be negligible. 

{\color{black}We use standard reduced quantities, namely reduced distances $l^* = l/\sigma$, reduced energies $E^*= E/\varepsilon$, and the reduced temperature is $T^*=k_{\text B}T/\varepsilon$, where $k_{\text B}$ is the Boltzmann constant, $T$ is an absolute temperature while $l$ and $E$ denote a distance and an energy, respectively.}  The temperature is kept at $T^*=1$.

We also define  the reduced density of the $i$th component as $\rho^*_k=\rho_k \sigma_k^3$, where $\rho=N_k/V$ is its number density, $N_k$ is the total number of ``atoms'' $k$  and V is the volume of the system.

\subsection{Simulation details}

Molecular dynamics simulations were carried out using the LAMMPS package \cite{50,51}.  The simulation box was a cuboid of reduced dimensions equal to {\color{black}$L_x^*=L_y^*=L_z^*=L^*=89$} along the axes $x$, $y$, and $z$, respectively. Standard periodic boundary conditions in the $x$ and $y$ directions were implemented. In the $z$-direction, two repulsive Lennard-Jones (12-6) walls were used to close the system.  The size $L_z^*$ was large enough to obtain two uniform bulk phases: $W$-rich and $O$-rich ones. A velocity-Verlet algorithm with a time step of  $\Delta t=0.002\tau$ was used for integrating the equations of motion, and a Nos\'e-Hoover thermostat with a time constant $\tau=10$ was used for regulating the system temperature. 

Each particle was constructed by placing the appropriate number of chains on the core surface.  First, a hairy particle was placed at the plane parallel to the $x$ and $y$ axes at $z = 0$. Then, equal numbers of molecules of the fluids $W$ and $O$ were inserted into the simulation box. During simulations, particles can freely rotate, translate and change their conformations.

The system was equilibrated for at least $10^7$ time steps until its total energy reached a constant level, at which it fluctuated around the mean value. The production runs were for at least $10^7$ time steps. To improve the accuracy, all averages are obtained from several independent simulations.

We carried the simulations for the following parameters of the particles: $\sigma_C^*= 6$, $m_C^*= 216$,  $M=20$, and $f=20$. Similar parameters were used by Tang et al.  \cite{38}.  To {\color{black}reduce} the number of parameters, we assumed that the majority of interactions between the species are repulsive. The considered interaction parameters are presented in the next section.

During simulations, we collected the density profiles of cores and segments, energies of interactions between all species, and total energy, and calculated various characteristics of the systems. Our goal was to estimate the position of the adsorbed particle, its orientation, and the shape of the polymer canopy. 

\subsection{Calculated observables}

To characterize the position of particles at the interface, we estimated the average distance of a core center from the phase boundary, i.e., the so-called displacement distance $h^*_c$. However, the shape of a segment cloud surrounding the core is characterized by the metrics generated by the gyration tensor \cite{52}

\begin{equation}\label{eq4}
G_{\alpha \beta}= \frac{1}{N'} \Big \langle \sum_{i=1}^{N'} {(r_{i,\alpha}-r_{0,\alpha})(r_{i,\beta}-r_{0,\beta})} \Big \rangle,
\end{equation}
where $r_{i,\alpha}$ and $r_{0,\alpha}$ are the $\alpha$ component ($\alpha, \beta=x, y, z$ ) of the $i$-th segment and the center of mass of the segment cloud, respectively. 

After diagonalization of the gyration tensor we obtain its eigenvalues $\lambda_i$ ($i = 1, 2, 3$),  ordered such that  $\lambda_1 \geqslant \lambda_2 \geqslant \lambda_3$.  Then, we calculate three invariants  $I_1=\lambda_1+\lambda_2+\lambda_3$, $I_2=\lambda_1\lambda_2+\lambda_1\lambda_3+\lambda_2\lambda_3$, and $I_3=\lambda_1\lambda_3\lambda_3$. Finally, these values are used  to define the shape descriptors of the segment cloud \cite{52}.
 
The radius of gyration of the segment cloud is given by
\begin{equation}\label{eq5}
R_g^2= I_1=
\frac{1}{N'} \Big \langle \sum_{i=1}^{N'} {\textbf{r}_{i0}^2 } \Big \rangle,
\end{equation}
where $\textbf{r}_{i0}=\textbf{r}_i-\textbf{r}_0$, and $\textbf{r}_i$ and $\textbf{r}_0$ are positions of the $i$-th 
segment and the center of mass, respectively, and $N' = fM$.

We resolve the vectors $\textbf{r}_{i0}$   into components parallel to the axes $x$, $y$, $z$, and calculate the corresponding radii of gyration labeled $R_{g \alpha}^2$ ($\alpha=x,y,z$). 

Moreover, we  define the relative shape anisotropy (asphericity)  $\kappa$, the prolateness $S^*$, and the acylindricity $C^*$ as
\begin{equation}\label{eq6}
\kappa=1-3 \langle I_2/I^2_1 \rangle,
\end{equation}
\begin{equation}\label{eq7}
S^*= \langle (3\lambda_1-I_1)(3\lambda_2-I_1)(3\lambda_3-I_1)/I^3_1\rangle,
\end{equation}
\begin{equation}\label{eq8}
C^*= \langle (\lambda_2-\lambda_3)/I_1 \rangle,
\end{equation}
where  $\langle ... \rangle$ denotes an average over all configurations.

The relative shape anisotropy takes on the values between 0 (perfectly spherical objects) and 1 (rigid rods). For a regular planar array, $\kappa= 0.25$ \cite{52}. The prolateness takes on the values between $-0.25$ and~2. For prolate shapes $S^*>0$ while negative values of $S^*$ correspond to oblate shapes. In a perfectly prolate segment cloud, the shorter axes of the ellipsoid are the same ($\lambda_1>\lambda_2=\lambda_3$). On the contrary, for perfectly oblate clouds, the longer axes are identical ($\lambda_1=\lambda_2>\lambda_3$).   The acylindricity is  $C^* \geqslant 0$ and for a perfect cylindrical symmetry $C^*=0$.  It should be stressed, however, that also a perfect sphere has a vanishing acylindricity.

\section{Results and discussion} \label{sec3}

The behavior of hairy particles at the liquid-liquid interface depends on the strengths of
interactions between all single entities: cores, segments, and liquid molecules. Various sets of interaction parameters were considered in simulations of hairy particles \cite{21,22}.  

\begin{figure}
\centering
\includegraphics[width=8.0cm]{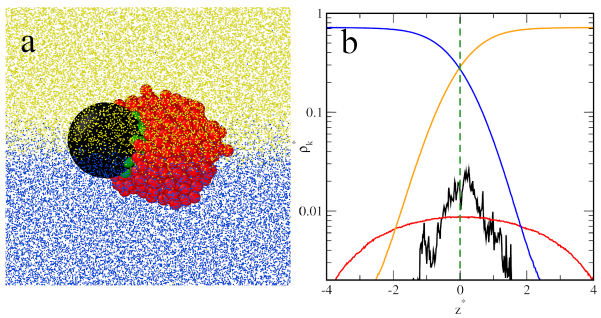}
\caption{(Colour online) Results for a hairy particle with purely repulsive ligands. (a) Exemplary configuration. The black sphere represents the core;  green spheres correspond to the segments bound to the core; red spheres represent other segments. Yellow and blue dots correspond to molecules $O$ and $W$, respectively. (b) Density profiles of the core (black line), the chain segments (red line), the profiles of liquids $O$ (orange line) and $W$ (navy line), $k=C,S,O,W$. The abscissa is scaled logarithmically. The green dashed line represents the location of the interface. }
\label{fig1}
\end{figure}

To model a phase boundary between immiscible liquids, we assume that {\color{black} interactions between molecules of the same liquids} are attractive, while  cross-interactions interactions are repulsive. We set $\varepsilon^*_{WW}=\varepsilon^*_{OO}=\varepsilon^*_{WO} =1$ \cite{38}. For attractive interactions between the same molecules ($OO$ and $WW$) $r^{(ij)}_{\text{cut}}=2.5\sigma_{(ij)}$, while repulsive interactions between different molecules ($OW$) $r^{(ij)}_{\text{cut}}=\sigma_{(ij)}$. Under the considered conditions, a very well-pronounced interface between these liquids is formed. {\color{black}The density profiles of both liquids drop rapidly to zero in the interphase region (see figure~\ref{fig1}b).}

We also introduce the following assumptions \cite{25,26,46}: (i) interactions of the cores with the remaining ``atoms'' are purely repulsive, (ii) {\color{black} interactions between the same ``atoms'' are attractive  ($\varepsilon^*_{ii}=1$).}

We consider the particles interacting (i) ``symmetrically''  and (ii) ``asymmetrically''  with both liquids. In the first case, ligands attract the molecules $O$ with the same strength as with the molecules $W$, $\varepsilon^*_{SO}=\varepsilon^*_{SW}$.   For the second type of particles interactions with molecules $W$ are assumed to be repulsive whereas the strength of attractive interactions with the molecules $O$ are attractive.   We carried out simulations for $\varepsilon^*_{SO}= 0.2, 0.4, 0.6, 0.8, 1.0, 1.2, 1.4, 1.6$.

{\color{black} Notice that different sets of energy parameters $\varepsilon^*_{SO}$ and $\varepsilon^*_{SW}$ can correspond to two types of real systems:  (i) different particles at the same interface (e.g., oil-water) or, conversely, (ii) the same particle  at the phase boundary between different liquids (more and less polar organic solvents).}

Our goal was to analyze how the mobility of the ligands and the strengths of interactions between the segments and both liquids influence the behavior of hairy particles at the interface, i. e., their shape position, and orientation. Examples of the equilibrium configurations are depicted using OVITO \cite{53aaa}. 

The behavior of the considered particles in bulk liquids depends on the strength of interactions between the ligands and liquid molecules {\color{black}\cite{8,10,11,12, 14, 20}}. If these interactions are repulsive or weakly attractive, the mobile ligands avoiding contact with the fluid form a cluster on the core's surface. In this way, the particle transforms into a kind of {\color{black}Janus-like particle} consisting of a naked part of the core and a cloud of segments.   As these interactions become stronger, the liquid molecules penetrate the polymer layer, so that the chains gradually move away from each other and a typical core-shell {\color{black}\cite{8,10,11,12, 14}} structure  is formed.
Below we discuss the behavior of such transformable particles at the liquid-liquid interface.

Figure \ref{fig1}a presents the exemplary configuration of a hairy particle with the segments repelling both liquids.  In this case, a Janus-like structure is formed.   Due to its shape, we refer to this configuration as  {\color{black} ``snowman-shaped particle'' \cite{r6}}. Such a particle adsorbs at the interface and ``isolates'' both liquids. In other words, the particle covers a part of the interface which leads to a decrease of a number of highly nonprofitable contacts between molecules $O$ and $W$.  The particle lies parallel to the interface. In figure \ref{fig1}b, the density profiles along the $z$-axis are presented. 
The core oscillated around the position at the interface ($z^*=0$). The segment profile has one relatively low and wide peak. The distribution of segments is symmetrical with respect to the phase boundary.

{\color{black} It should be emphasized  here that the attachment energy of Janus particles is  several times higher  than that for corresponding homogeneous particles \cite{4,33,34}. 
In this case, the particles exhibit amphiphilic properties. The localization of a particle at the interface gives an additional energetic profit associated with different interactions of particular parts of the particle with both liquids. Janus particles have a greater capability to stabilize the emulsions.}

\begin{figure}
\centering
\includegraphics[width=14.0cm]{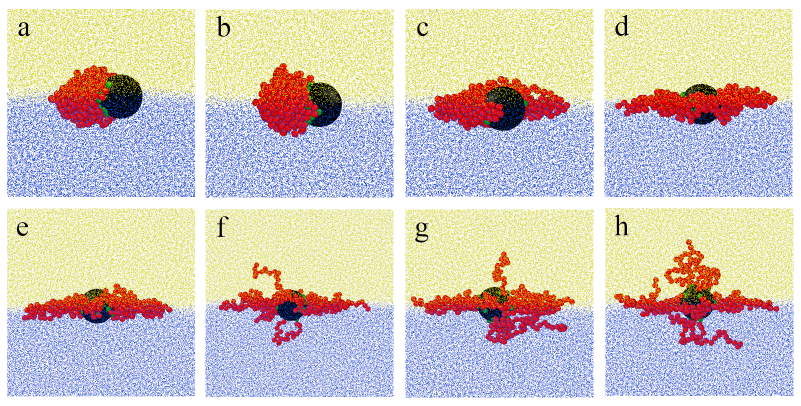}
\caption{(Colour online) Equilibrium configurations for symmetrically interacting hairy particles (side views), for different values of energy parameters $\varepsilon^*_{SO}= \varepsilon^*_{WO}$: (a) 0.2, (b) 0.4, (c) 0.6, (d) 0.8, (e) 1.0, (f) 1.2, (g) 1.4, and (h) 1.6. The black sphere represents the core;  green spheres correspond to segments bound to the core; red spheres represent other segments. Yellow and blue dots correspond to molecules $O$ and $W$, respectively.}
\label{fig2}
\end{figure}

\begin{figure}
\centering
\includegraphics[width=14.0cm]{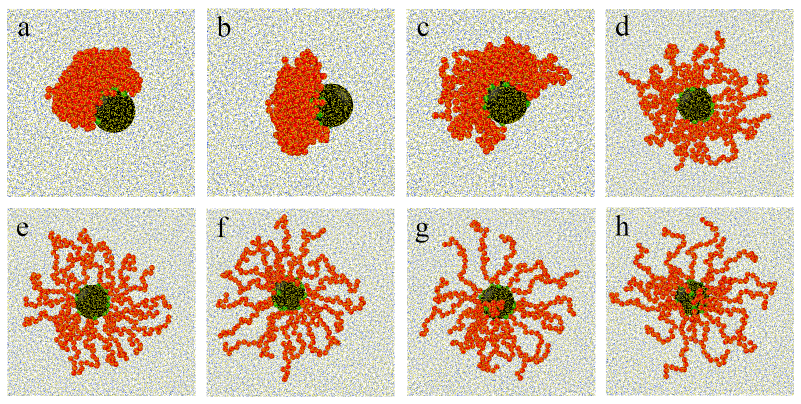}
\caption{(Colour online) Equilibrium configurations for symmetrically interacting hairy particles (views from above) for different values of energy parameters $\varepsilon^*_{SO}= \varepsilon^*_{WO}$: (a) 0.2, (b) 0.4, (c) 0.6, (d) 0.8, (e) 1.0, (f) 1.2, (g) 1.4, and (h) 1.6. The black sphere represents the core;  green spheres correspond to the segments bound to the core; red spheres represent other segments. Yellow and blue dots correspond to molecules $O$ and $W$, respectively.}
\label{fig3}
\end{figure}

Now, we  turn to the discussion of the behavior of particles with symmetrical attractive interactions with the liquids.  We have collected the configurations of such particles in figure \ref{fig2} (side view) and figure~\ref{fig3} (top view). As previously, for weak attractive interactions, {\color{black}``snowman-shaped Janus-like particles'' } are observed (parts a-c). When the strength of these interactions increases, the cloud of segments gradually ``spills'' at the liquid surface. Liquid molecules penetrate the brush and the chains straighten up and they move away from each other. In this way, a Saturn-like configuration is formed (see parts d and e). Further increasing $\varepsilon^*_{SO}$ causes partial pulling of the chains into the depths of both liquids. The energy gain resulting from the interaction of the chains with the liquids outweighs the gain associated with the reduction of the contact surface between the liquids (parts g and h).

\begin{figure}
\centering
\includegraphics[width=12.0cm]{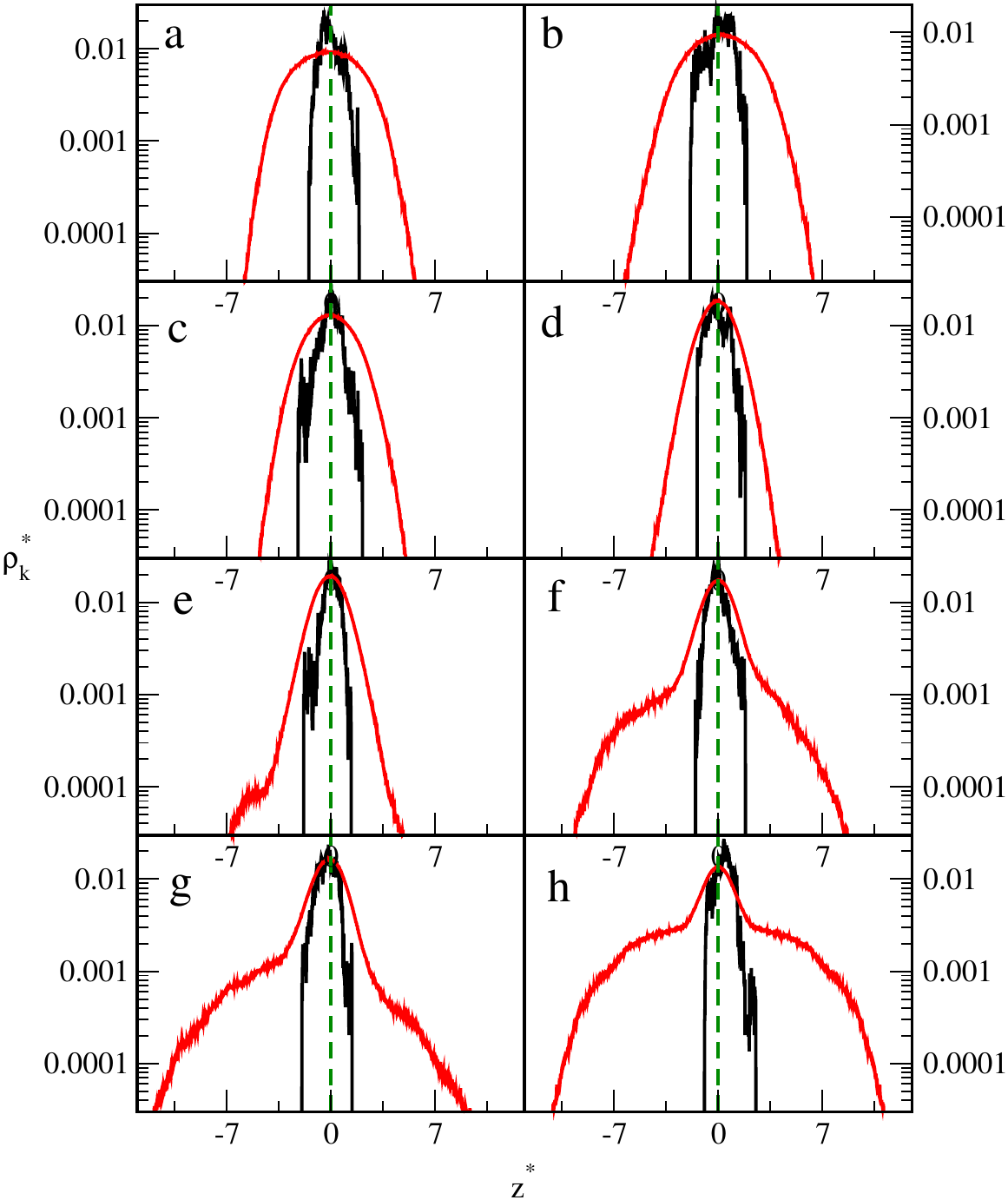}
\caption{(Colour online) Density profiles of the core ($k=C$, black lines) and the chain segments ($k=S$, red lines) for symmetrically interacting hairy particles and different values of energy parameters $\varepsilon^*_{SO}= \varepsilon^*_{WO}$: (a) 0.2, (b) 0.4, (c) 0.6, (d) 0.8, (e) 1.0, (f) 1.2, (g) 1.4, and (h) 1.6. The green dashed lines represent the location of the interface. The abscissas are scaled logarithmically.}
\label{fig4}
\end{figure}

To quantify these observations, we calculated the corresponding density profiles (see figure \ref{fig4}). In all cases, one narrow and symmetrical peak at $z^*=0$ was obtained for the core. However, the segment density profiles significantly vary as the parameter $\varepsilon^*_{SO}$ increases. Obviously, the segment profiles are symmetrical and have  maxima at $z^*=0$. Initially, the segment distributions have typical Gaussian shapes and become narrower and slightly higher (parts a-d). This reflects the spilling of the polymers on the interface. For stronger interactions, the profiles become wider and their shape considerably changes (parts e-h). The segment density sharply decreases around $z^*=0$, then this decline is slower and finally rapidly falls to zero. 
 
\begin{figure}
\centering
\includegraphics[width=14.0cm]{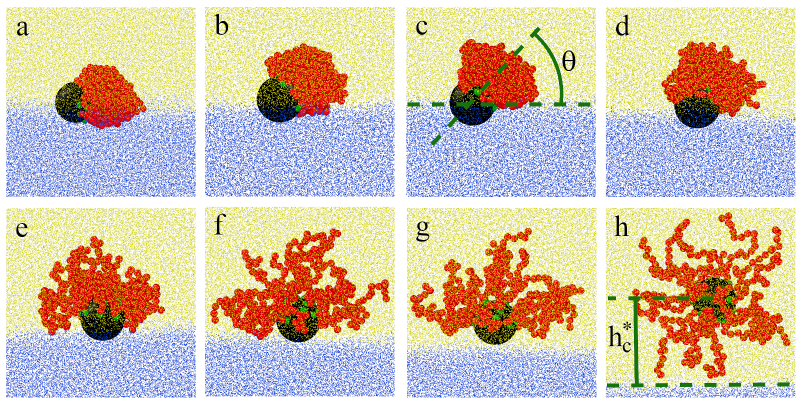}
\caption{(Colour online) Equilibrium configurations for asymmetrically interacting hairy particles (side views) and different values of the energy parameter $\varepsilon^*_{SO}$: (a) 0.2, (b) 0.4, (c) 0.6, (d) 0.8, (e) 1.0, (f) 1.2, (g) 1.4, and (h) 1.6. Interactions between segments and molecules $W$ are repulsive. The black sphere represents the core;  green spheres correspond to segments bound to the core; red spheres represent other segments. Yellow and blue dots correspond to molecules $O$ and $W$, respectively. The angle, $\theta$ characterizing the particle orientation, and the displacement distance, $h^*_C$ are shown in (c) and (h), respectively.}
\label{fig5}
\end{figure}

A situation of particles with asymmetrical interactions at the interface is much different from the previous case.  Examples of their equilibrium configurations are presented in figure \ref{fig5}.  The ligands repel molecules $W$ and attract molecules $O$. At the interface, the particles form the above-mentioned Janus-like {\color{black} snowman-shaped structures}. For weak interactions with the liquid $O$, the ligands form a compact ``drop'' of segments (parts a-d). As the interactions become stronger,  the chains gradually straighten in the preferred liquid ($O$), avoiding the phase boundary (parts e-g).   On the one side of the core, an unfurled plume is formed. Moreover, with an increasing value of $\varepsilon^*_{SO}$, the core shifts towards the liquid $O$, and for sufficiently strong interactions, the particle leaves the interface (h). 

\begin{figure}
\centering
\includegraphics[width=12.0cm]{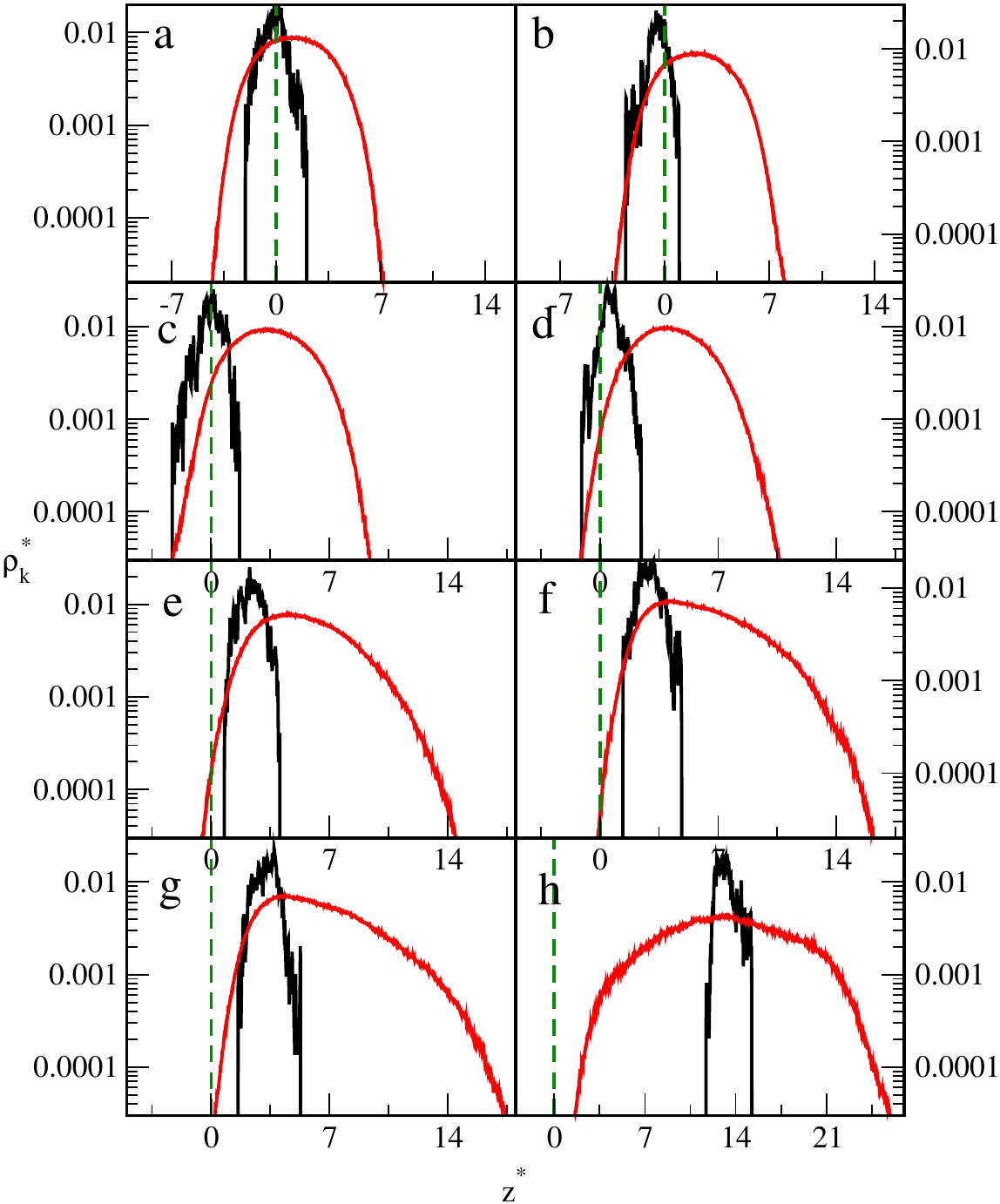}
\caption{(Colour online) Density profiles of the core ($k=C$, black lines) and the chain segments ($k=S$, red lines) for asymmetrically interacting hairy particles and different values of the energy parameter $\varepsilon^*_{SO}$: (a) 0.2, (b) 0.4, (c) 0.6, (d) 0.8, (e) 1.0, (f) 1.2, (g) 1.4, and (h) 1.6. Interactions between segments and molecules $W$ are repulsive. The abscissas are scaled logarithmically. The green dashed lines represent the location of the interface.}
\label{fig6}
\end{figure}

The analysis of the density profiles plotted in figure \ref{fig6} confirms these insights. For weak interactions (parts a-c),  the core density has a maximum at $z^*=0$. This means that the core is pinned to the interface. A further increase in $\varepsilon^*_{SO}$ causes its partial entering into phase $O$ (parts d-g), and a maximum is shifted towards the liquid $O$ (unpinned configurations). In the case of weak interactions, the segment density profiles are symmetrical peaks located at $z^*>0$ (from the $O$-liquid side). On the contrary, for stronger interactions, these peaks are asymmetrical, extended towards the liquid $O$. However, when the particle goes to the bulk phase, the segment distribution has a form of a symmetrical and wide peak (part h).

\begin{figure}
\centering
\includegraphics[width=12.0cm]{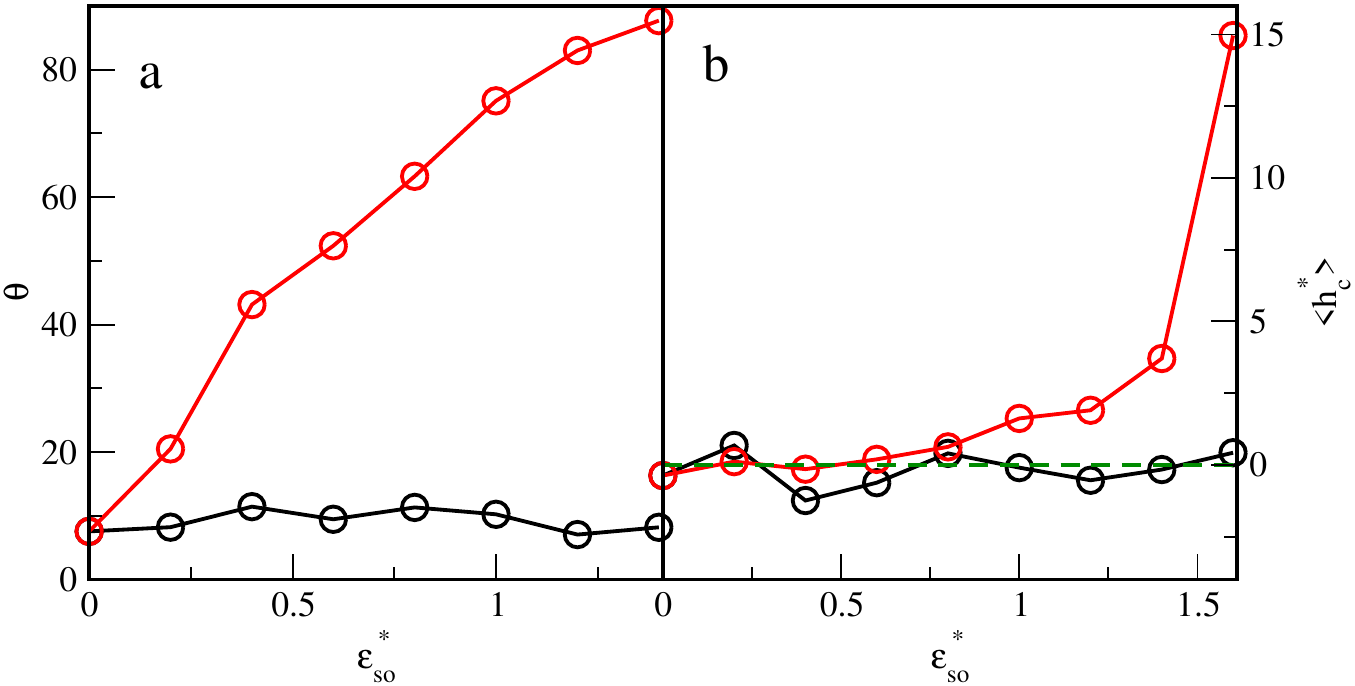}
\caption{(Colour online) (a) Orientation angle, $\theta$ and (b) the average position of the core (displacement distance), $h^*_C$, for symmetrical interactions particles (black circles) and asymmetrically interacting particles with the segments repulsive for $W$ molecules (red circles) as functions of the energy parameter $\varepsilon^*_{SO}$. The green dashed lines represent the location of the interface. }
\label{fig7}
\end{figure}

To characterize the orientation of particles, we use the angle between the $xy$-plane and the line connecting the mass centers of the clouds of segments and the center of the core (the angle $\theta$ is shown in figure \ref{fig5}c) {\color{black}\cite{4, 33,34}}. Another parameter characterizing a particle configuration at the interface is the average displacement depth, i.e., the distance of the center of the core from the phase boundary{\color{black}\cite{4, 33,34}} (see figure \ref{fig5}h).  These quantities are presented in figure \ref{fig7} for particles interacting symmetrically (black lines) and asymmetrically (red lines) as a function of $\varepsilon^*_{SO}$.   The results obtained for repulsive ligands are plotted at $\varepsilon^*_{SO}=0$. We treat these interactions as infinitely weak attractive ones.  

In the case of symmetrical interactions (figure \ref{fig7}a), the particle lies parallel to the phase boundary to occupy as much area on it as possible (compare figure \ref{fig2}). However, for asymmetrical interactions with the liquid, the particle can be differently oriented with respect to the phase boundary. As the attractive interactions with one liquid become stronger, the particle gradually changes its orientation, from tilted (figure \ref{fig5} a--g) to normal (figure \ref{fig5}h). For symmetrically interacting particles, the $h^*_c$ oscillates around zero (figure \ref{fig7}b). When the interactions are asymmetrical and comparable or stronger than the interactions with the preferred liquid, the displacement depth quickly increases. Such a behavior of hairy particles results from a complex interplay between the reduction of the area of the liquid-liquid interface caused by its adsorption and interactions between all species. This is altered by the reconfiguration of a polymer canopy built of mobile ligands. {\color{black}Qualitatively similar behavior was reported for Janus ellipsoids and Janus dumbbells at liquid-liquid interfaces \cite{4}}.

\begin{figure}
\centering
\includegraphics[width=7.0cm]{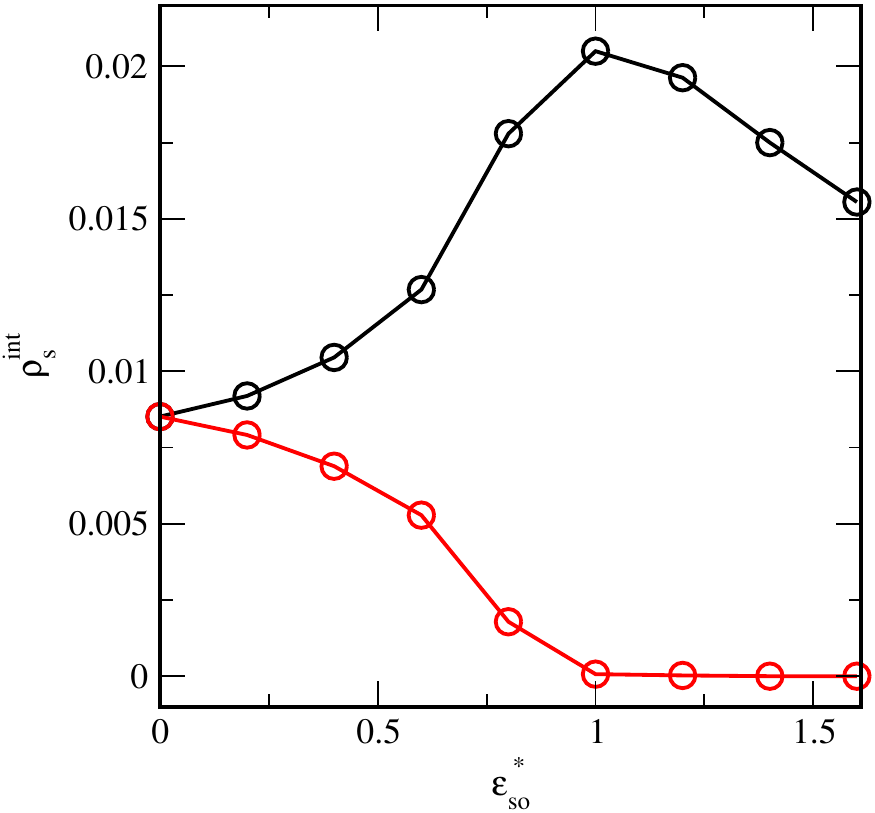}
\caption{(Colour online) The density of segments in the surface layer  $z^*_C \in {<}-0.5,0.5{>}$  for symmetrically interacting particles (black circles) and asymmetrically interacting particles with the segments repulsive for $W$ molecules (red circles) as functions of the energy parameter $\varepsilon^*_{SO}$.}
\label{fig8}
\end{figure}

We now turn  our attention to the question of how the shape transformations at the interface are affected by an increase in the strength of  $SO$ interactions.
Firstly, we analyze the average density of segments in the surface layer  $z^*_C \in {<}-0.5,0.5{>}$ (see figure \ref{fig8}). The segments located here block the access to the interface. In the case of symmetrical interactions, this density increases to a maximum at $\varepsilon^*_{SO}=1$ and then decreases. Indeed, for $\varepsilon^*_{SO}>1$, a part of the chains starts to unfold in the direction of the liquid $O$. Thus, the effect of the covering of the interface is the strongest for $\varepsilon^*_{SO}=1$.  However, for asymmetrically interacting particles, the surface density quickly decreases, and for $\varepsilon^*_{SO}=1$ it reaches zero. With increasing $\varepsilon^*_{SO}$, such particles change their orientations and adsorption height ($h^*_z$). The segments ``escape'' from the surface. 

\begin{figure}
\centering
\includegraphics[width=11.0cm]{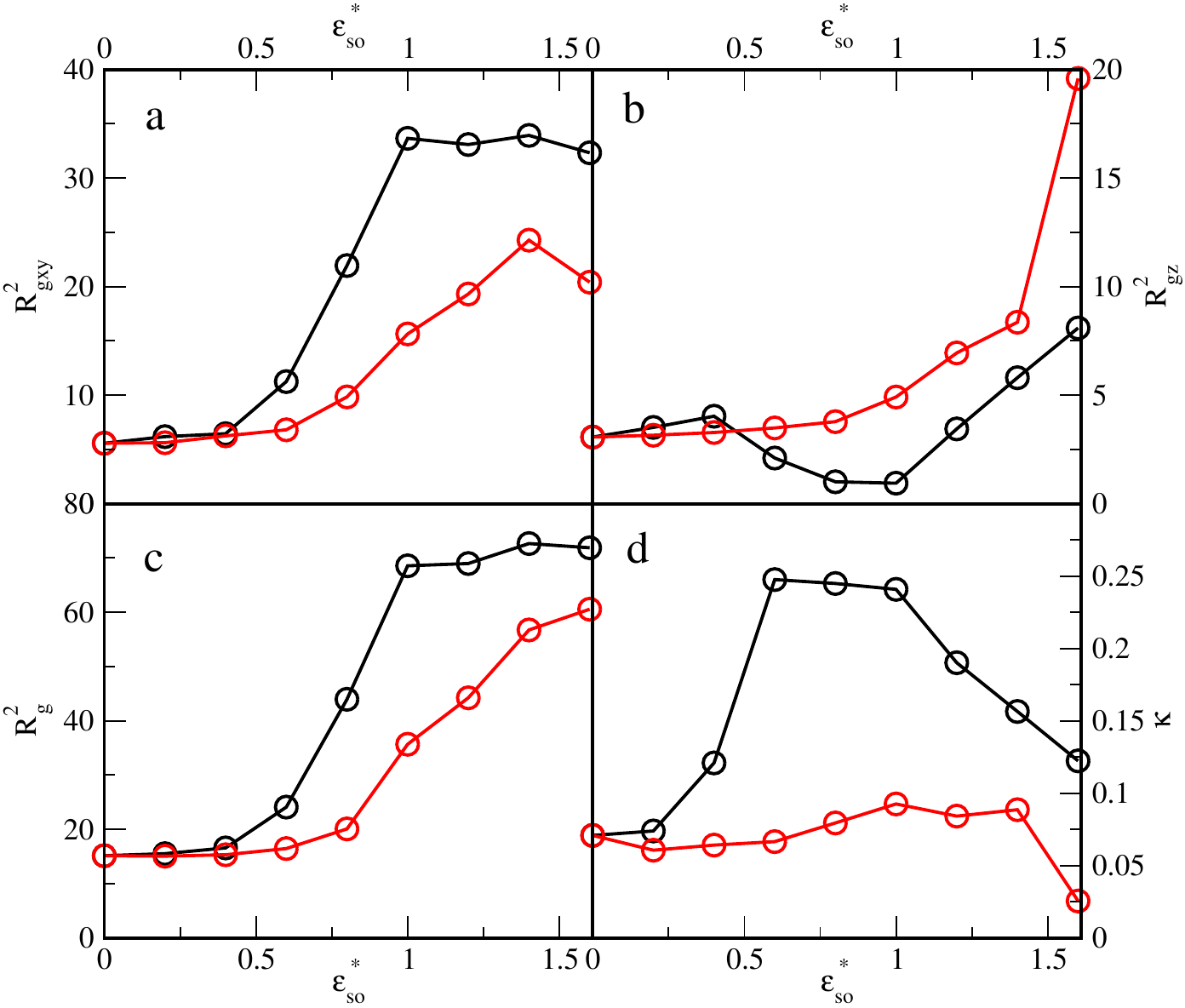}
\caption{(Colour online) Shape metrics for symmetrically interacting particles (black circles) and asymmetrically interacting particles with the segments repulsive for $W$ molecules (red circles) as functions of the energy parameter $\varepsilon^*_{SO}$. (a) The average squared radius of gyration in directions parallel to the liquid-liquid interface; (b) the $z$th component of the squared radius of gyration; (c) The total radius of gyrations, and (d) the relative shape anisotropy. }
\label{fig9}
\end{figure}

Further corroboration for the reconfigurable nature of the hairy particles is offered by an analysis of their geometrical characteristics. For this purpose, we employ  the shape metrics computed starting from the gyration tensor. In figure \ref{fig9}, we plot the total squared radius of gyration, its components and the relative shape anisotropy as a function of $\varepsilon^*_{SO}$.  In all cases, the components in $x$ and $y$ directions are almost the same. To improve the statistics, we calculated their averages, $R^2_{gxy}$. The averaged squared radii of gyration in directions parallel to the interface are presented in part a. The $z$-th components of $R^2_g$ are shown in part b, while part c depicts the total squared radii of gyration.  Figure \ref{fig9}d shows how the shape anisotropy changes with an increase in the strength of $SO$ interactions.

We begin with the analysis of the curves obtained for symmetrically interacting particles. As $\varepsilon^*_{SO}$  increases, the squared radius of gyration $R^2_{gxy}$ rises and reaches the plateau at $\varepsilon^*_{SO}=1$ (Saturn-like configurations). However, the $z$-component initially slightly increases, then decreases to the minimum at  $\varepsilon^*_{SO} \approx 1$  and quickly increases. This well reflects the effect of flattening the cloud of segments on the phase boundary. Obviously, the later rise of the $R^2_z$ corresponds to the penetration of the chains into the preferred liquid. The total squared radius of gyration varies in the same way as the squared radius of gyration in the directions $x$ and $y$. The asphericity of symmetrically interacting particles rapidly increases from 0.07 (repulsive interactions) to approximately 0.25 for  $0.6 \leqslant \varepsilon^*_{SO} \leqslant 1.0$ and then quickly decreases. This result confirms the formation of quasi-two-dimensional structures of segments for $\varepsilon^*_{SO}=1$. With a further increase in $\varepsilon^*_{SO}$, the polymer canopy becomes more and more spherical. 

Other relations are found for asymmetrically interacting particles. As the energy parameter increases to $\varepsilon^*_{SO}=1.4$, the squared radii of gyration $R^2_{gxy}$, the $R^2_{gz}$ and $R^2_g$ monotonously increase.  Initially, the chains remain coiled but for stronger interactions the polymer layer considerably swells.  When the particle detaches from the interface ($\varepsilon^*_{SO}=1.6$), the $R^2_{gxy}$ slightly decreases, while $R^2_{gz}$ sharply rises. It is interesting that the relative shape anisotropy is relatively low ($\kappa < 0.1$) and remains almost unchanged for all particles attached to the interface. For $\varepsilon^*_{SO}=$, however, the $\kappa$ drops to a value close to the limit value, 0.  Thus, the relative shape anisotropy identified all the observed structures as almost spherical ones.

\begin{figure}
\centering
\includegraphics[width=12.0cm]{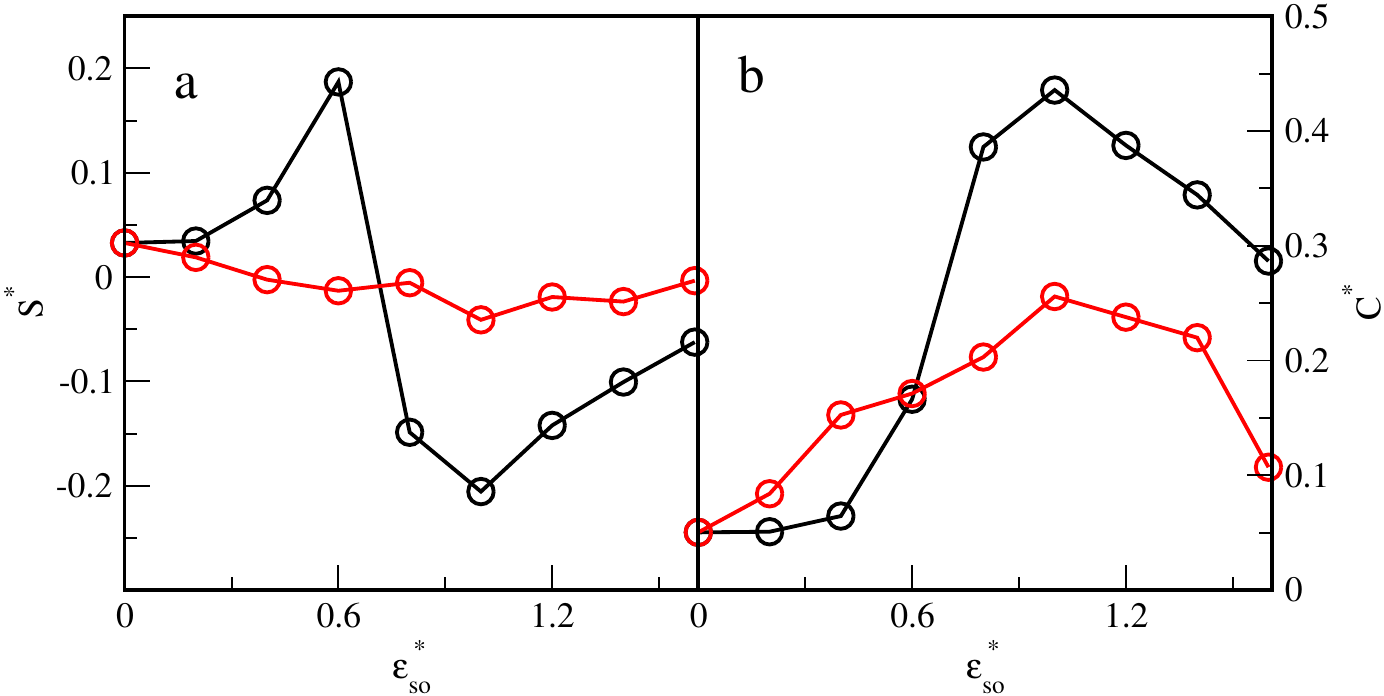}
\caption{(Colour online) The prolateness (a) and the acylindricity (b) for symmetrically interacting particles (black circles) and asymmetrically interacting particles with the segments repulsive for $W$ molecules (red circles) as functions of the energy parameter $\varepsilon^*_{SO}$.}
\label{fig10}
\end{figure}

In figure \ref{fig10} we report the prolateness $S^*$ and acylindricity $C^*$ as a function of the energy parameter $\varepsilon^*_{SO}$.  In the case of symmetrical interactions,  for weak interactions ($\varepsilon^*_{SO} \leqslant 0.6$), the cloud of segments becomes more and more prolate while it is oblate at greater values of $\varepsilon^*_{SO}$. A qualitative and rapid change is observed between values $\varepsilon^*_{SO}=0$ and $\varepsilon^*_{SO}=0.8$. For $\varepsilon^*_{SO}=1$, the cloud is almost perfectly oblate, and the prolateness reaches the minimum value $S^*=-0.2$.  These changes, well reflect the transformations from {\color{black} snowman-shaped particles} to Saturn-like structures and later ``standing up'' of chains from the surface,  as shown in figures \ref{fig1}, \ref{fig2}, and \ref{fig3}.  This is also visible at the plot $C^*$ vs. $\varepsilon^*_{SO}$. The acylindricity increases from 0.05 to a maximum value  $C^*=0.45$ at $\varepsilon^*_{SO}=1$. In a Saturn-like particle, the polymer canopy has the shape of a hockey puck and deviates considerably from a perfect cylinder of elongated shape. For stronger interactions, acylindricity slowly decreases.

For asymmetrically interacting particles considered here, the prolateness is always close to zero. However, for $\varepsilon^*_{SO} \leqslant 0.2$,  the $S^*$ is positive but for stronger interactions $S^*$ is negative. The prolateness reaches a minimum value for $\varepsilon^*_{SO}=1$. However, the acylindricity increases from 0.05 to 0.25 (for $\varepsilon^*_{SO}=1$) and drops to 0.11. In this case, the relative shape anisotropy is also close to zero. These two metrics classified our clouds as spherical.

{\color{black}Finally, it should be stressed that in the case of particles with permanently anchored ligands, their shape changes are less spectacular \cite{37, r7}. Such hairy particles also retain a core-shell structure  at the liquid-liquid interface. Depending on the interactions with the liquids, their coronas take on the shape of increasingly elongated ellipsoids \cite{r7}. However, the formation of Janus-like structures  has not been reported so far. }

{\color{black}Our study shows how the hairy particles with mobile ligands change the shape when they are trapped at a liquid-liquid interface.} 

\section{Conclusions} \label{sec4}

We used molecular dynamics simulations to study individual spherical particles modified with mobile ligands at the interface between partially miscible Lennard-Jones liquids ($O/W$). To limit the number of parameters, we assumed that almost all interactions are repulsive except the interactions between the same molecules of liquids and the ligand-liquid interactions. 

We considered two types of particles: particles symmetrically and asymmetrically interacting with both liquids. In the first case, interactions of chains with molecules of both liquids are the same. For the second group of particles, interactions of ligands with the liquid $W$ are repulsive while their interactions change from repulsive to strongly attractive. We found that particles which in the same way interact with both liquids are always pinned to the interface but a particle shape is strongly affected by the energy parameter $\varepsilon^*_{SO}$. 

Adsorption of symmetrically interacting particles results only from the coverage of a part of the phase boundary, which leads to the reduction of energetically unfavorable contacts between particles of different liquids. Such particles are always pinned to the interface but a particle shape is strongly affected by the energy parameter $\varepsilon^*_{SO}$. For weak interactions, particles take on the form of {\color{black} snowman-shaped Janus-like particles} consisting of a part of the bare core and a ``drop'' of segments and orient parallel to the phase boundary. {\color{black}As interactions with liquid become stronger, a Saturn-like structure is formed.} A further increase in $\varepsilon^*_{SO}$ causes the chains to be drawn deeper into the preferred phase and particles take on the form of a core with a plume.

If interactions of particles with liquids are different, the surface activity of particles can be additionally altered. In this case, we observe various equilibrium configurations. The center of the core can be located either at the interface (pinned configuration) or slightly shifted into a preferred liquid (unpinned configuration). An increase of the energy parameter $\varepsilon^*_{SO}$ causes the shape transformation of particles. For weak interactions, Janus-like structures pinned to the interface are found. Initially, the {\color{black} particles} lie at the interface but with increasing strength of interactions the particles align themselves at an angle to the surface.
{\color{black}  The particle} orientation gradually changes from parallel, through tilted to perpendicular. {\color{black}This is similar to the behavior of Janus ellipsoids and Janus dumbbells at the liquid-liquid interface \cite{4}.} In the case of stronger interactions, the chains straighten in the preferred liquid and ``escape'' from the interface. Currently, the core slightly shifts towards the interior of one of the phases.   On the one side of the core, an unfurled plume is visible. 

We also discuss how the interactions with liquids affect the metric properties characterizing the shape of segment clouds, such as the radius of gyration, its components in the Cartesian coordinates, the relative shape anisotropy,  the prolateness, and acylindricity. Our analysis shows that the chosen quantities well reflect the observed structures. 

{\color{black}As shown in numerous studies \cite{2,4,38,r7}, particles with permanently anchored ligands also change shapes at the liquid-liquid interface. The particles with symmetrical interactions with both liquids deform into oblate ellipsoids to maximize the occluded area at the interface. In the case of different interactions with liquids, the polymer coronas  become asymmetrical. However, such particles do not form Janus-like and Saturn-like structures. We conclude that ligand mobility increases the capability of hairy particles to change the internal morphology at the interface.  This strengthens their surface activity, i.e., their tendency to stay at the interface. Due to this, such particles can be effectively used to stabilize the emulsions.}

We believe that our approach contains  salient physical characteristics to capture the mechanisms of {\color{black}rearrangement} of hairy particles trapped at liquid-liquid interfaces. 
{\color{black}The study proves that ligand mobility strongly affects the behavior of an individual hairy particle at a liquid-liquid interface.  Our work may be a starting point for investigation of the impact of ligand mobility on the adsorption of hairy particles at the interface,  their self-assembly and  their capability to stabilize Pickering emulsions.}


\bibliographystyle{cmpj}
\bibliography{bibliography}

\ukrainianpart

\title{Зміни форми волосистої частинки з рухливими лігандами на межі розділу ``рідина-рідина''}
\author[Т. Сташевскі, М. Борувко]{Т. Сташевскі, М. Борувко}
\address{Факультет теоретичної фімії, Інститут хімічних наук, Університет ім. Марії Склодовської-Кюрі, Люблін, Польща}

\makeukrtitle

\begin{abstract}
	Ми досліджуємо перегрупування однієї волосистої частинки на межі поділу ``рідина-рідина'' з використанням огрубленої молекулярної динаміки. Розглядаються частинки з однаковою (симетричні взаємодії) та різною (асиметричні взаємодії) спорідненістю до рідин. Показано, як рухливість ліганду впливає на поведінку волосистої частинки на межі поділу ``рідина-рідина''. З'ясовано, що така волосиста частинка може приймати різні форми на межі розділу: наприклад, подібного на Януса сніговика, що складається зі скуп\-чен\-ня сегментів і оголеної частини ядра; структур, подібних до Сатурна; ядра з широким ``шлейфом'' з одного боку. Конфігурація частинки на межі розділу характеризується відстанню вертикального зміщення та орієнтацією частинки відносно межі розділу фаз. Вибрані дескриптори використовуються для характеристики форми сегментної хмари. Виявлено, що форму частинки та її локалізацію на межі розділу можна визначити шляхом регулювання її взаємодії з рідинами.
	\keywords волосисті частинки, збагачені частинками шари, молекулярна динаміка
\end{abstract}

 \end{document}